\documentclass{article}

\usepackage{arxiv}

\usepackage[utf8]{inputenc} 
\usepackage[T1]{fontenc}    
\usepackage{hyperref}       
\usepackage{url}            
\usepackage{booktabs}       
\usepackage{amsfonts}       
\usepackage{nicefrac}       
\usepackage{microtype}      
\usepackage{lipsum}
\usepackage{authblk}
\usepackage{graphicx}
\usepackage{amsmath}
\usepackage{comment}
\usepackage{xcolor}
\usepackage{adjustbox}
\usepackage{booktabs}
\usepackage{subcaption}
\usepackage{float}
\usepackage{algorithmic}
\usepackage{caption}
\usepackage{cite}
\usepackage{amsmath,amssymb,amsfonts}
\usepackage{algorithmic}
\usepackage{graphicx}
\usepackage{textcomp}
\usepackage{xcolor}
\usepackage{siunitx}
\usepackage{balance}
\usepackage{url}

\title{A Data Pilot-Aided Temporal Convolutional Network for Channel Estimation in IEEE 802.11p Vehicle-to-Vehicle Communications}

\author[1,2,3]{\textbf{S.~A.~Ngorima}}
\author[1]{\textbf{A.~S.~J.~Helberg}}
\author[1,2,3]{\textbf{M.~H.~Davel}}

\affil[1]{Faculty of Engineering, North-West University, South Africa}
\affil[2]{Centre for Artificial Intelligence Research, South Africa}
\affil[3]{National Institute for Theoretical and Computational Sciences, South Africa}
\affil[ ]{\texttt{aldringorima@gmail.com}}

\date{}
\date{\vspace{-2em}}

\begin{document}

\maketitle

\begingroup
\renewcommand{\thefootnote}{}
\footnote{This work is a preprint of a published paper by the same name\cite{Ngorima2024}. The authenticated version is available online at \url{https://www.satnac.org.za/proceedings} pages 356-361.}
\endgroup

\begin{abstract}
In modern communication systems, having an accurate channel estimator is crucial. However, when there is mobility, it becomes difficult to estimate the channel and the pilot signals, which are used for channel estimation, become insufficient. In this paper, we introduce the use of Temporal Convolutional Networks (TCNs) with data pilot-aided (DPA) channel estimation and temporal averaging (TA) to estimate vehicle-to-vehicle same direction with Wall (VTV-SDWW) channels. The TCN-DPA-TA estimator showed an improvement in Bit Error Rate (BER) performance of up to 1 order of magnitude. Furthermore, the BER performance of the TCN-DPA without TA also improved by up to 0.7 magnitude compared to the best classical estimator.
\end{abstract}


\section{Introduction}

Beyond 5G networks and the modern Wi-Fi standard used for vehicular communication systems are expected to deliver high data speeds and low latency~\cite{10314524}. However, mobility in wireless channels can cause doubly selective fading, where the channel changes both in the temporal and frequency domains. This is mainly due to multipath propagation, which results in varying delays, Doppler shifts, and attenuation as users move. It is important to accurately estimate the channel because doubly selective fading affects receiver processes such as equalisation, demodulation, and decoding, which directly influences system performance. For reliable communication in mobile scenarios, accurate channel estimation algorithms are vital to handle the doubly selective fading nature of wireless channels.

Pilot signals are usually used to estimate the channel, but in mobility scenarios these pilots become insufficient~\cite{gizzini2021temporal}. To address this, a data pilot-aided (DPA) estimation scheme is generally used, in which data subcarriers are remapped and used for channel estimation~\cite{7421323}. Several classical methods are then built upon DPA. This includes $(1)$ the spectral temporal averaging (STA) scheme~\cite{5982455}, where the averaging is performed in the time and frequency domains, $(2)$ the constructed data pilot (CDP) scheme~\cite{6646362} in which the previous symbols are used as preambles to estimate the current symbol, and $(3)$ time domain reliability test frequency domain interpolation (TRFI)~\cite{6957832}, where a reliability test is performed on the DPA estimated channel to obtain reliable subcarriers. The reliable channel estimates are interpolated at the unreliable subcarriers. 

The process of demapping the data subcarriers in DPA to obtain an initial channel estimate is heavily influenced by the noise level and the accuracy of the previously estimated channel~\cite{pan2021channel}. As the previous data symbols are utilised as the preamble to obtain the current symbol, the error in the DPA increases as we progress from one symbol to the next across the frame, leading to performance degradation in vehicular environments. Given that DPA is the initial step in most classical methods, this error is also present in classical channel estimation methods.  

In this work, we propose the use of TCN-based estimators to reduce DPA errors and improve performance in vehicular channel environments. Using a TCN for initial channel estimation, we can achieve more accurate results, which can then be utilised effectively by the DPA method to estimate the current channel state as a post-processing step. In addition, we also test the effectiveness of temporal averaging (TA) in further improving the channel estimate.

The remainder of the paper is structured as follows: Section~\ref{background} provides background information on the IEEE 802.11p standard investigated in this paper. Section~\ref{review} offers a review of current channel estimation schemes, while Section~\ref{TCN} introduces the TCN-based estimator. In Section~\ref{results}, the results are presented and analysed. Section~\ref{conclusion} concludes the paper.
\section{Background}
\label{background}
This section summarises the IEEE 802.11p standard specification and examines channel estimation system models and techniques in the context of IEEE 802.11p.
\subsection{ System Model}
\begin{figure}[ht] 
    \centering
    \includegraphics[width=0.5\linewidth]{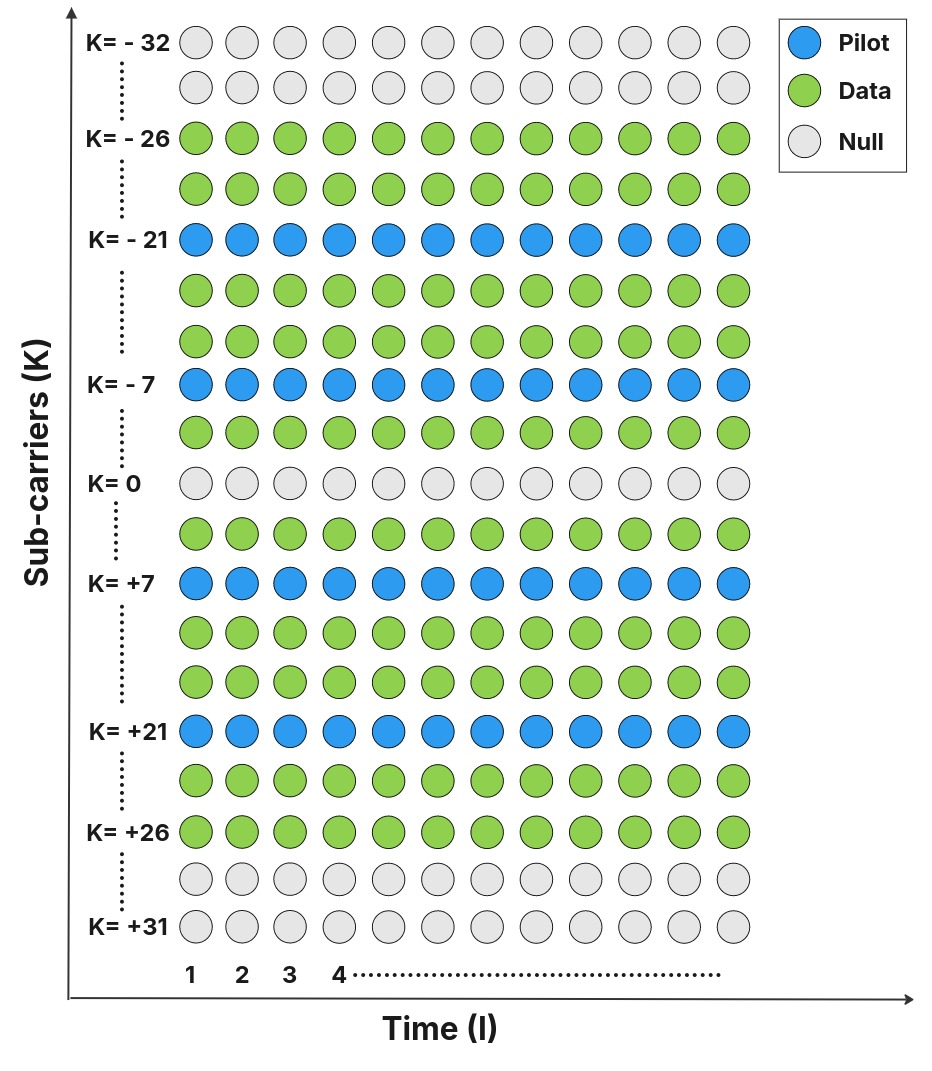}
    \caption{IEEE 802.11p subcarrier arrangement \cite{gizzini2020deep}.}
    \label{fig:subcarr}
\end{figure}

The IEEE 802.11p protocol uses Orthogonal Frequency Division Multiplexing (OFDM) to facilitate data transmission across radio channels. OFDM divides the available bandwidth (10MHz) into subcarrier frequencies, allowing multiple signals to be sent at the same time~\cite{abdelgader2014physical}. Pilot signals on designated subcarriers track channel conditions to enhance data decoding for efficient communication. Only 52 of the 64 available subcarriers are used for data transmission and pilot symbols, while the remaining subcarriers are used for guard bands and direct current (DC) offset. Fig.~\ref{fig:subcarr} shows how these subcarrier indices are arranged in the IEEE 802.11p frame.
The four pilot subcarriers are positioned at indices
\{\num{-21}, \num{-7}, 7, 21\}, while the remaining 48 indices are designated for data.

In this study, we assume perfect synchronisation and consider a frame structure that begins with two extended preambles, followed by \textit{I} data symbols. The received OFDM symbol  $y_i[k]$ at time instant \textit{i} and subcarrier \textit{k} can be represented as in~\cite{gizzini2020deep}:

\begin{equation*} y_{i}[k]=h_{i}[k]x_{i}[k]+n_{i}[k],\tag{1} \label{rec_signal}\end{equation*}
where $x_i[k]$ is the transmitted constellation symbol, $h_i[k]$ is the channel frequency response, $n_i[k]$ is the noise added to the \textit{i-th} OFDM symbol on the  \textit{k-th }subcarrier. 

\subsection{IEEE 802.11p Channel Estimation Methods}
\label{review}
In this section, we provide an overview of the IEEE 802.11p channel estimation methods. All the methods discussed will be evaluated against our proposed approach.

\subsubsection{LS Estimation}
The least squares (LS) estimation method is the widely accepted channel estimation technique in wireless communications to estimate the channel response. It is also a standard method used in IEEE 802.11p. LS utilises the two received preambles as follows~\cite{gizzini2021temporal}:
\begin{equation*} \hat { {h} }_{\text {LS}}[k] = \frac { {y} ^{(p)}_{1}[k] + {y} ^{(p)}_{2}[k]}{2 {p} [k]}, \tag{2}\label{eqn:ls}\end{equation*}
where ${y} ^{(p)}_{1}[k]$ and ${y} ^{(p)}_{2}[k]$ represent the received preambles located at the beginning of each frame at subcarrier \textit{k} and $p[k]$ is the predefined preamble that was transmitted.  

\subsubsection{DPA Estimation}
\label{DPA}
In vehicular communications, DPA is considered the initial channel estimation method in vehicular communications because it is assumed that there is a high correlation between the successive OFDM symbols received~\cite{gizzini2020deep}.
In DPA, the LS estimation at the preambles (as described by (\ref{eqn:ls})) is used to obtain the initial channel estimate; the subsequent channel estimate is obtained by equalising the received symbols with the initial channel estimate as described by (\ref{eqn:yeq}). Equalised symbols are demapped to the closest constellation as described by (\ref{eqn:conse}) and the final DPA channel estimate is updated as expressed in (\ref{eqn:dpa_update}). According to~\cite{gizzini2021temporal}, the remaining OFDM symbols are equalised using the previously estimated channel:
\begin{equation*}
y_{i}^{\text{eq}}[k] = \frac{y_{i}[k]}{\hat{h}_{i-1}^{\text{DPA}}[k]}, \tag{3}
\end{equation*}

\begin{equation}
\hat{h}_{0}^{\text{DPA}}[k] = \hat{h}_{\text {LS}}[k], 
\tag{4}
\label{eqn:yeq}
\end{equation}
As outlined in~\cite{gizzini2020deep}, the data subcarrier $y_{i}^{eq}[k]$ is then demapped and mapped to the nearest constellation point using the respective modulation order to determine its closest constellation point $d_i[k]$ which is given by 
\begin{equation*}d_{i}[k]=\mathfrak{R}\left(\frac{y_{i}[k]}{ \hat{h}_{i-1}^{\text{DPA}}[k]}\right), \tag{5}\label{eqn:conse}\end{equation*}  
where $\mathfrak{R}\left(.\right)$ is a mapping operation to obtain the closest constellation point. The initial channel estimate $\hat{h}_{0}^{\text{DPA}}[k]$ is derived from the LS estimation, expressed as
\begin{equation}
    \hat{h}_{0}^{\text{DPA}}[k]=\hat{h}_{\text {LS}}[k],
    \label{eqn:dls}
    \tag{6}
\end{equation}
and the final DPA channel estimate is subsequently updated as
\begin{equation*} \hat{h}_{i}^{\text{DPA}}[k]=\frac{y_{i}[k]}{d_{i}[k]}.\label{eqn:dpa_update}\tag{7}\end{equation*}

\subsubsection{STA Estimation}
The STA method continuously updates the channel estimate using data symbols, leveraging time and frequency correlation across OFDM symbols. Following~\cite{gizzini2020deep} the STA channel estimate is obtained by performing all the steps of the DPA procedure, followed by frequency domain averaging of the estimates for all selected subcarriers as described below:
\begin{equation*} \hat{h}_{i}^{\text{FD}}[k]=\sum_{\lambda=-\beta}^{\beta}\omega_{\lambda}\hat{h}_{i}^{\text{DPA}}[k+\lambda],\label{eqn:sta_first}\tag{8}\end{equation*}
\begin{equation}
    ~\omega _{\lambda } = \frac {1}{2\beta +1},
    \tag{9}
    \label{eqn:sta_sec}
\end{equation}
where $2\beta + 1$ denotes the number of subcarriers being averaged. 
Finally, time averaging is performed to calculate the final STA channel estimate as
\begin{equation*}\hat{h}_{i}^{\text{STA}}[k]=\left(1-\frac{1}{\alpha}\right)\hat{h}_{i-1}^{\text{STA}}[k]+\frac{1}{\alpha}\hat{h}_{i}^{\text{FD}}[k]\tag{10}\label{eqn:sta_final}\end{equation*}
where $\alpha$ is a filtering parameter that varies depending on the characteristics of the channels.

\subsubsection{CDP Estimation}
The CDP method also relies on the high correlation of OFDM symbols. CDP performs all DPA procedures. Then, as described in~\cite{gizzini2020deep}, the previously received symbol is equalised using both the current DPA estimate, $\hat{h}_{i}^{\text{DPA}}[k]$, and the previous CDP estimate, $\hat{h}_{i-1}^{\text{CDP}}[k]$ as

\begin{equation}
{y}_{i-1}^{\text{eq}\prime}[k] = \frac{y_{i-1}[k]}{\hat{h}_{i}^{\text{DPA}}[k]},
\tag{11}
\label{eqn:cdp_eq_prime}
\end{equation}
\begin{equation}
{y}_{i-1}^{\text{eq}\prime\prime}[k] = \frac{y_{i-1}[k]}{\hat{h}_{i-1}^{\text{CDP}}[k]},
\tag{12}
\label{eqn:cdp_eq_double_prime}
\end{equation}

The equalised symbols ${y}_{i-1}^{\text{eq}\prime}[k]$ and ${y}_{i-1}^{\text{eq}\prime\prime}[k]$ are then demapped to their closest constellation ${d}_{i-1}^\prime[k]$ and ${d}_{i-1}^{\prime\prime}[k]$ using the mapping operation $\mathfrak{R}\left(.\right)$. As outlined in~\cite{gizzini2020deep}, the final CDP channel estimate $\hat{h}_{i}^{\text{CDP}}[k]$ is obtained as follows 
\begin{equation*}\hat{h}_{i}^{\text{CDP}}[k]=\begin{cases}\hat{h}_{i-1}^{\text{CDP}}[k], & {d}_{i-1}^\prime[k]\neq{d}_{i-1}^{\prime\prime}[k],\\\hat{h}_{i}^{\text{DPA}}[k], & {d}_{i-1}^\prime[k]= {d}_{i-1}^{\prime\prime}[k]\end{cases}\tag{13}\label{eqn:cdp}\end{equation*}
If ${d}_{i-1}^\prime[k]\neq{d}_{i-1}^{\prime\prime}[k]$, it indicates that $\hat{h}_{i}^{\text{DPA}}[k]$ is not reliable and therefore the previous channel estimate should be used instead. 

\subsubsection{TRFI Estimation}
The TFRI method is a channel estimation technique designed to enhance the performance of CDP. The steps involved in the TFRI process are similar to the CDP estimator processes. The main difference is that, instead of the comparison step described by (\ref{eqn:cdp}), TFRI uses the demapped symbols ${d}_{i-1}^\prime[k]$ and ${d}_{i-1}^{\prime\prime}[k]$ calculated after the processes expressed by (\ref{eqn:cdp_eq_prime}), and (\ref{eqn:cdp_eq_double_prime}) to create a reliable subcarrier set and an unreliable set. Then it uses the reliable set to interpolate the channel at unreliable subcarriers based on a reliable test as detailed by the authors in~\cite{gizzini2020deep}. 

\section{Proposed TCN-DPA Estimator}
\label{TCN}
TCNs are designed to handle sequential data, whether it is single-channel or multiple-channel sequential data, making them ideal for capturing temporal dependencies in the received signal. Using 1-dimensional convolutional layers, TCNs can efficiently process sequential data, identifying patterns that classical methods may miss. This section outlines the architecture of the TCN and the proposed TCN-based estimator used in this work. 
\subsection{Temporal Convolutional Networks (TCNs)}
\begin{figure}[H]
    \centering
    \begin{subfigure}[b]{0.49\linewidth} 
        \includegraphics[width=\linewidth]{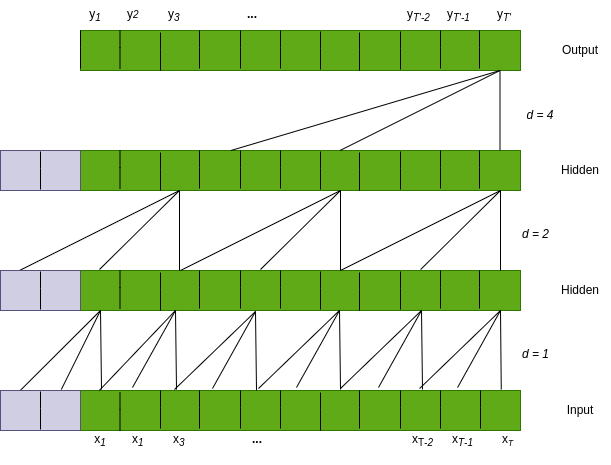}
        \caption{Causal dilated convolutions.}
        \label{fig:Causal}
    \end{subfigure}
    \begin{subfigure}[b]{0.48\linewidth} 
        \includegraphics[width=\linewidth]{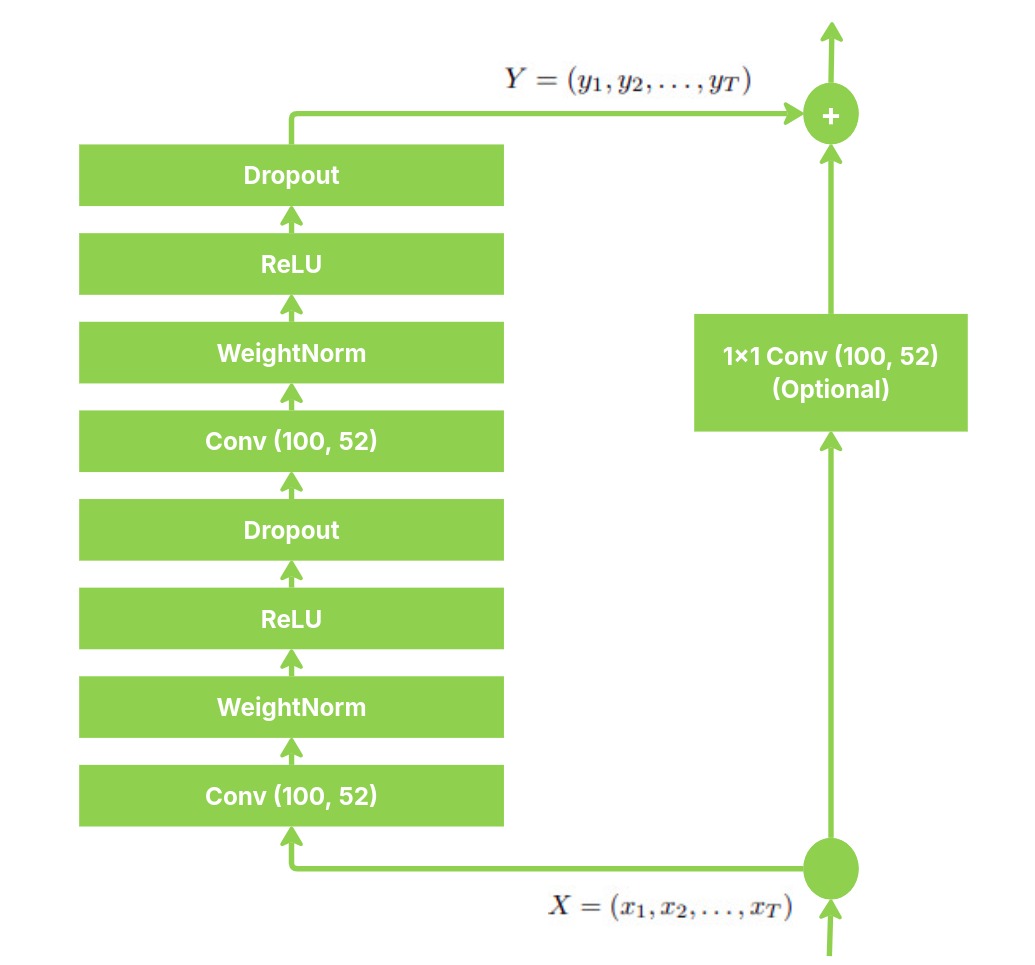}
        \caption{TCN Residual block adapted from~\cite{bai2018empirical}.}
        \label{fig:RB}
    \end{subfigure}
    \caption{TCN architectural elements. (a) Causal dilated convolutions and (b) TCN residual block with an optional 1x1 residual connection that can be added when the input and output have different dimensions.}
    \label{fig:combined}
\end{figure}

Convolutional Neural Networks (CNNs), introduced by LeCun~\cite{lecun1995convolutional}, are key models in the development of neural networks (NNs). CNNs use convolutional and pooling layers to extract local features and reduce dimensionality. 
TCNs were developed from CNNs, specifically to handle sequential data. TCNs integrate specialised components such as causal convolution, dilated convolution, and residual connections, as shown in Fig.~\ref{fig:combined}~\cite{bai2018empirical}, making them effective in capturing temporal dependencies within the input data.

A TCN uses one-dimensional convolution kernels (kernels that move in only one direction, for example from left to right) to analyse sequential data in the temporal domain, as shown in Fig.~\ref{fig:Causal}. Suppose that we have an input sequence $\{x_0$,\ldots,$x_T\}$ and our goal is to forecast the corresponding outputs $\{y_0$,\ldots,$y_T\}$ at each time step. The main constraint is that when predicting the output $y_t$ at any given time t, we can only use the inputs that have been observed up to that point: $\{x_0$,\ldots, $x_t\}$. To ensure that the output is of the same length as the input, zero padding of length (kernel size - 1) is applied to the hidden layers. A TCN uses causal convolutions to ensure that there is no future information leakage.
For causality to be true, $y_T$ should depend only on $\{x_0$,\ldots,$x_T\}$ and not on any future input beyond time $T$.

A crucial component of a TCN is the residual block~\cite{bai2018empirical}. It incorporates a branch that takes the input sequence and applies a series of transformations, denoted as $F$. When the outputs have dimensions different from the input, a residual branch is present where the outputs are then combined with the original input $x$ through an activation function, as: 
\begin{equation}
\text{Output} = \text{Activation}(x + F(x)) \tag{14}\label{eqn:activation}
\end{equation}
The residual connection allows the layers to focus on capturing the desired mapping instead of learning the entire transformation from input to output. The residual block typically contains dilated causal convolution layers and non-linearity layers. One commonly used non-linearity is the Rectified Linear Unit (ReLU), which enhances network sparsity and helps prevent overfitting by setting a portion of the neuron's output to zero. The ReLU function also addresses the issue of gradient vanishing during backpropagation. Its mathematical definition is given as:
\begin{equation*}
{\text{f}}(x,w,b) = \max \left( {0,{w^T}x + b} \right) \tag{15}
\end{equation*}
where $x$ represents the input to the neuron, $w$ denotes the weight parameter, and $b$ signifies the bias term.
In the residual block, the dropout layer is applied during training after each convolutional layer to prevent overfitting and improve the model's generalisation capabilities.
\subsection{Dataset Generation}
This section discusses the characteristics of the generated dataset, the vehicular channel model used, and the structure of the dataset for use by a TCN.
\subsubsection{Channel Model}
The vehicular channel model investigated in this work is specified in \cite{4350097}. In this study, we investigate the vehicle-to-vehicle same direction with wall (VTV-SDWW) tapped delay line (TDL) vehicular channel model. This model is used for communication between two vehicles moving in the same direction with a central wall separating them while keeping a distance of 300-400 metres between them.
A VTV-SDWW TDL mobility scenario is used in which vehicles have a velocity of 100 km/h and a Doppler shift of $f_d$ = 550 Hz. 16QAM modulation is used.

\begin{figure}[ht]
    \centering
    \includegraphics[width=\linewidth]{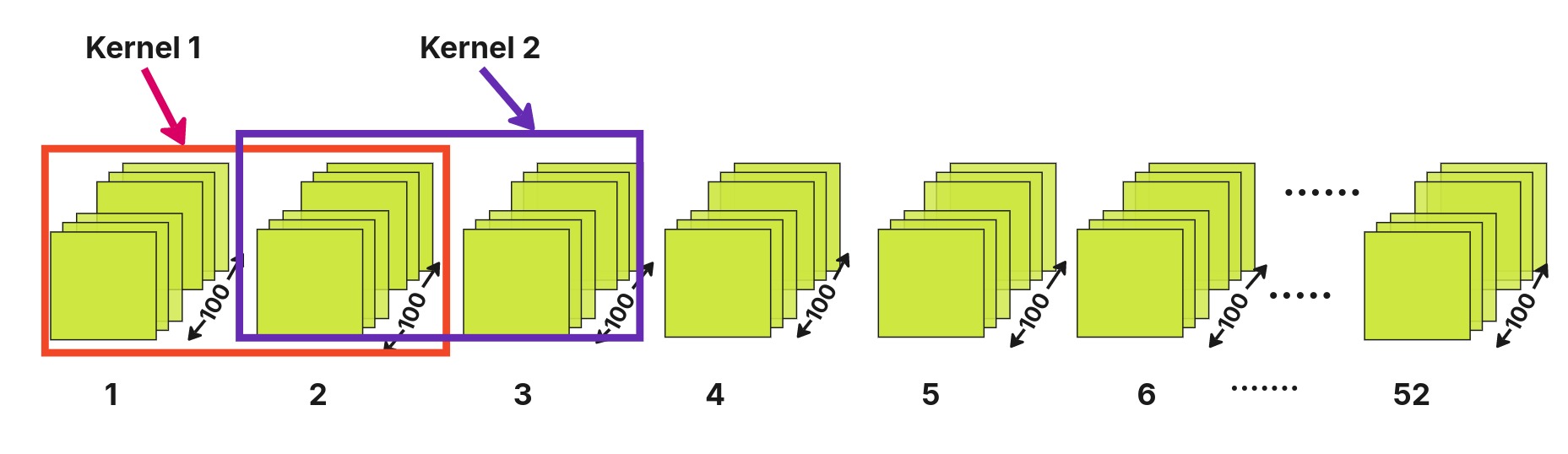}
    \caption{TCN operation with an input sequence of 52 subcarriers and 100 channels of interleaved real and imaginary parts of the complex data.}
    \label{kernel}
\end{figure}

Our dataset consists of 18,000 time-specific samples or frames. Each sample includes a sent and received version of 50 OFDM symbols. Each OFDM symbol consists of a complex value for each of the 52 active subcarriers, where 48 are data subcarriers and 4 are pilot subcarriers. To prepare the data for the NN, we separate the complex-valued received 16 QAM symbol per resource element and concatenate them as shown in Fig.~\ref{kernel}, such that the values are interleaved per subcarrier and time slot. This appears as if the number of symbols received per carrier has doubled. Each input sample is now a matrix of real values with dimensions of $52 \times 100$ and each output is a matrix of real values with dimensions of $48 \times 100$ where the pilot signals are excluded. These samples are grouped into batches. 

The input tensor for the TCN is a 3D tensor with dimensions $(N, C, T)$ where $N$ represents the batch size, $C$ represents the number of channels (100 separated and interleaved real and imaginary parts of the complex data) and $T$ represents the number of `time steps' (in this case, not actual time, but the 52 subcarriers are modelled as if sequential). In our case, with a batch size of 128, the input tensor shape is $(128, 100, 52)$. A kernel size of 2 indicates the length of the convolutional filter. This filter processes two consecutive time steps across 100 channels to calculate weighted sums of the input values it overlaps.

The NN is trained at the 40 dB SNR level. This is based on the hypothesis of the authors in \cite{gizzini2020joint, gizzini2021temporal, gizzini2020deep} that training a NN on a high SNR dataset facilitates the NN to learn the actual channel state and this is important for the NN. 

\subsection{TCN-DPA estimator}
This section describes the proposed TCN-DPA estimator where the initial input to the TCN is the DPA estimate in the frequency domain (the subcarriers). The DPA process is performed after this TCN process. The previous TCN output, \(\hat{h}_{i-1}^{\text{TCN}}[k]\), is used by the DPA process to equalise the current received symbol. The equalised symbol is then mapped to the closest constellation point, acquiring the demapped symbol as expressed in (\ref{eqn:TCN}). The final channel estimate \(\hat{h}_{i}^{\text{TCN}}[k]\) is updated as: 
\begin{equation*}d_{i}^{\text{TCN}}[k]=\mathfrak{R}\left(\frac{y_{i}[k]}{ \hat{h}_{i-1}^{\text{TCN}}[k]}\right),\label{eqn:dem_tcn}\tag{16}\end{equation*} 
\begin{equation}
    \hat{h}_{0}^{\text{TCN}}[k]=\hat{h}_{\text {DPA}}[k],\tag{17}\label{eqn:TCN}
\end{equation}
\begin{equation*}\hat{h}_{i}^{\text{TCN}}[k]=\frac{{y}_{i}[k]}{d_{i}^{\text{TCN}}[k]}\tag{18}\label{eqn:final_TCN_est}\end{equation*}
\begin{figure}[ht]
    \centering
    \includegraphics[width=0.7\linewidth]{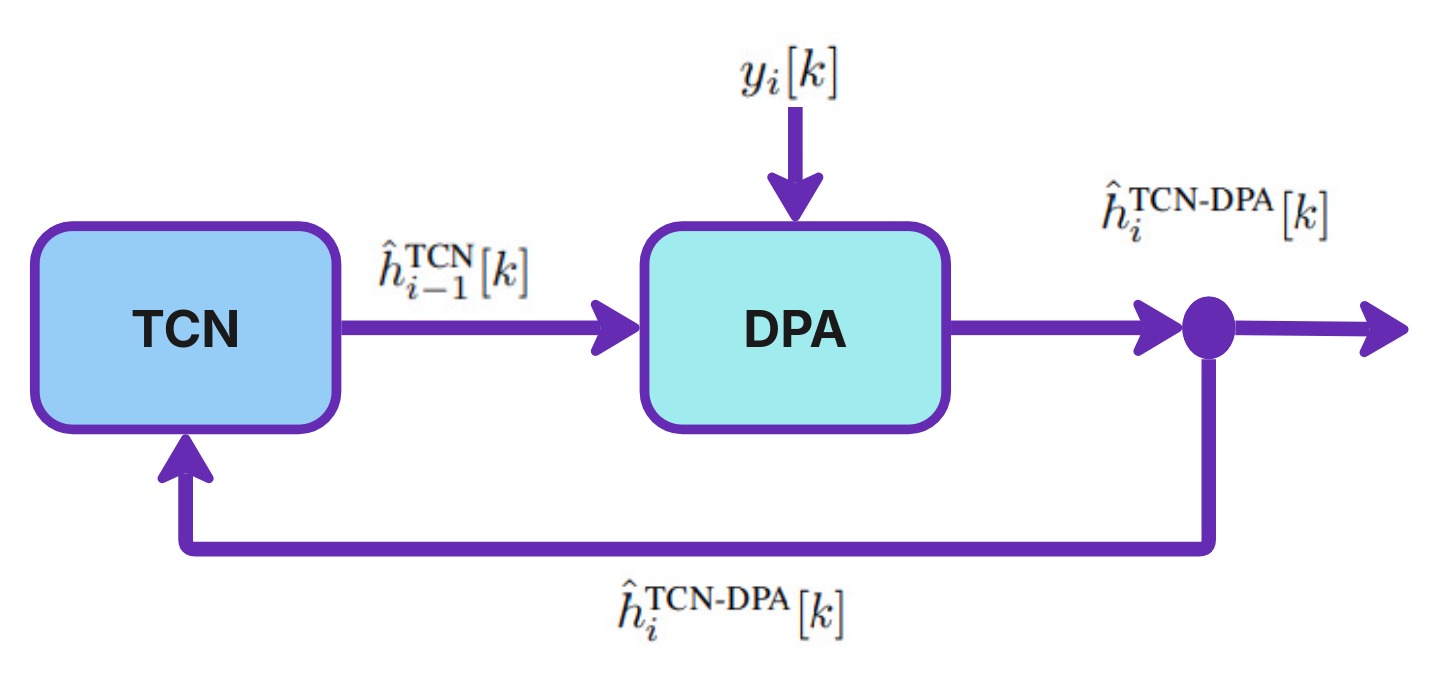}
    \caption{TCN-DPA estimator process showing how the TCN output is used for the DPA process.}
    \label{fig:error_comp}
\end{figure}
Fig.~\ref{fig:error_comp} shows this iterative process. The next section gives a description of the TA process and how it can be added to the TCN-DPA method.     
\subsubsection{TA Processing}

\begin{figure}[ht]
    \centering
    \includegraphics[width=0.8\linewidth]{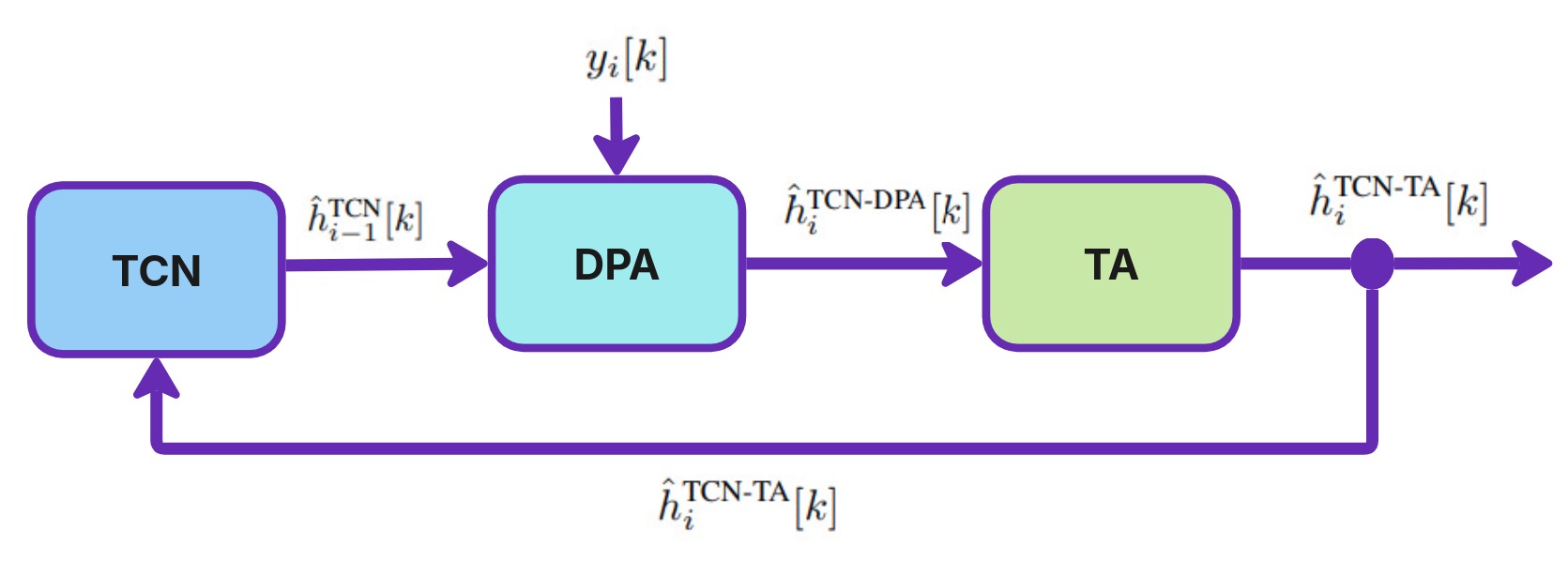}
    \caption{TCN-DPA-TA estimator process showing the TCN output as input to the DPA and the TA processes and how the TCN-DPA-TA output is used to update the final channel estimate.}
    \label{fig:h_dpa_ta}
\end{figure}
Temporal averaging (TA) is a method for noise reduction in which the noise reduction ratio can be determined analytically. As stated in~\cite{gizzini2021temporal}, TA can iteratively decrease the Additive White Gaussian Noise (AWGN) power within a frame. In that study, an $\alpha$ value of 2 yielded commendable results. For simplicity, we also use an $\alpha$ of 2 in this work. The complete TCN-DPA-TA is shown in Fig.~\ref{fig:h_dpa_ta}.

The TA process is applied to the estimated TCN-DPA channel $\hat{h}_{i}^{\text{TCN-DPA}}[k]$ so that the final channel estimate $\hat{h}_{i}^{\text{TCN-TA}}[k]$ is obtained as: 
\begin{equation*}\hat{h}_{i}^{\text{TCN-TA}}[k]=\left(1-\frac{1}{\alpha}\right)\hat{h}_{i-1}^{\text{TCN-TA}}[k]+\frac{1}{\alpha}\hat{h}_{i}^{\text{TCN-DPA}}[k]\label{eqn:h_dpa_ta}\tag{19}\end{equation*}

where $\alpha$ represents the weight size used to calculate the averages. A higher value of $\alpha$ means that more symbols are used. 

\subsubsection{TCN Hyperparameter Tuning}
\label{sec:hyper}
In this work, we use Bayesian optimisation (BO) to find the best hyperparameter combination to improve model performance. BO constructs a probabilistic model, typically a Gaussian process, to approximate the function based on observed data. Gaussian processes are preferred as surrogate models due to their ability to provide both mean predictions and uncertainty estimates for input parameters~\cite{8743779}.
We optimise the learning rate, number of layers, kernel size, dropout, and StepLR parameters (step size and gamma). For hyperparameter tuning, we use Optuna~\cite{akiba2019optuna} and Weights \& Biases (WandB)\footnote{\url{https://wandb.ai/}}. Optuna, a BO framework, uses a tree-structured parzen estimator (TPE) to maintain a probabilistic model linking hyperparameters to outcomes. This approach enables efficient exploration of the hyperparameter space, refining it sequentially to improve the probability of reaching the global optimum in fewer trials. WandB is used for logging and monitoring experimental results.

We conduct 80 trials to explore the hyperparameter space comprehensively. Each trial runs for 200 epochs to allow the models to converge and for the effects of hyperparameter changes to become evident. The validation loss is used as an indicator of convergence and the optimal hyperparameters are selected based on the lowest validation loss observed during these trials. A search space for the hyperparameters is defined as shown in Table \ref{tab:hyperparameters}. The best hyperparameter combination is obtained and is also shown in the same table.

\begin{table}[ht!]
    \begin{center}
    \renewcommand{\arraystretch}{1.5}
    \caption{Optimal TCN Hyperparameters.}
    \begin{tabular}{|c|c|c|}
    \hline
    \textbf{Hyperparameter} & \textbf{Search Space} & \textbf{Optimal Value} \\
    \hline
    Learning Rate & $1 \times 10^{-5}$ to $1 \times 10^{-2}$ & 0.003 \\
    \hline
    Number of Layers & 1 to 5 & 4 \\
    \hline
    Kernel Size & 2 to 5 & 2 \\
    \hline
    Dropout & $10^{-5}$ to 0.5 & 0.01 \\
    \hline
    StepLR Step Size & 10 to 50 & 17 \\
    \hline
    StepLR Gamma & 0.5 to 1 & 0.8 \\
    \hline
    Epochs & 0 to 200 & 100 \\
    \hline
    \end{tabular}
    \label{tab:hyperparameters}
    \end{center}
\end{table}

\section{Analysis and Results}
\label{results}

The TCN-DPA architecture consists of a 4-layer TCN residual block with a kernel size of 2 and layer dilations of (1, 2). 
12,000 frames are used for training, 4,000 for validation, and 2,000 are used for testing.
\begin{figure}[H]
    \centering
    \begin{subfigure}[b]{0.49\linewidth} 
        \includegraphics[width=\linewidth]{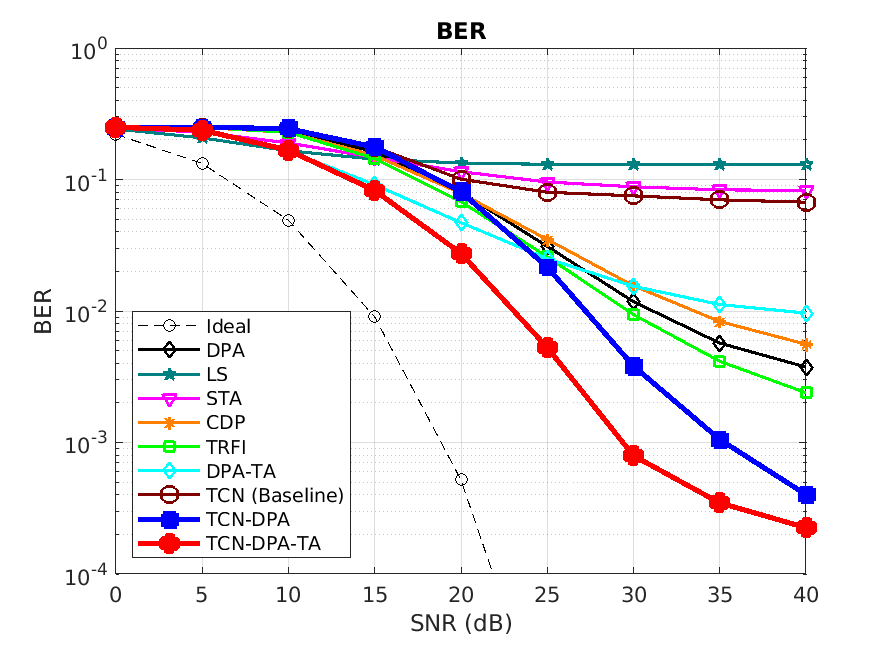}
        \caption{BER simulation results of various estimators.}
        \label{fig:TCN-no-TA-BER}
    \end{subfigure}
    \hfill
    \begin{subfigure}[b]{0.49\linewidth} 
        \includegraphics[width=\linewidth]{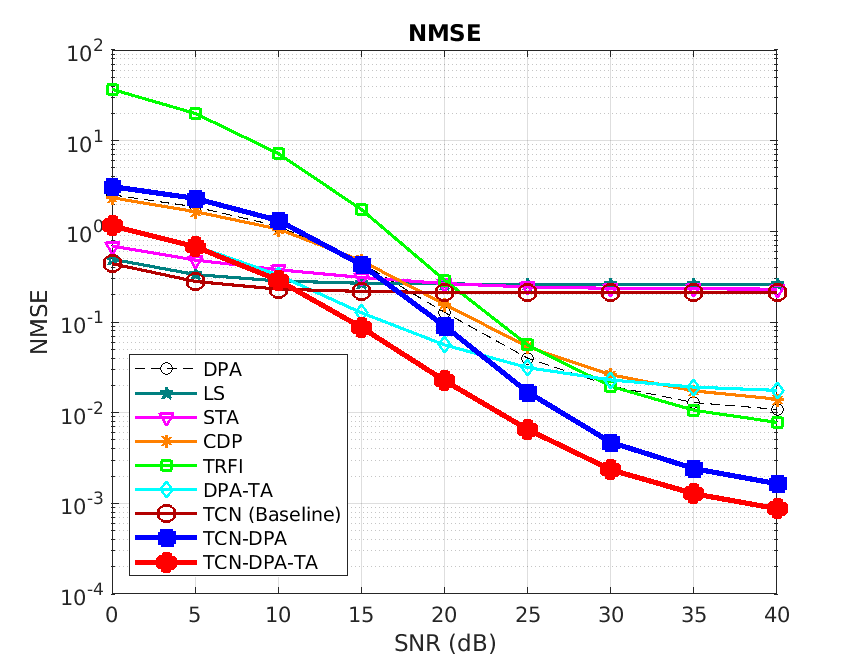}
        \caption{NMSE simulation results of various estimators.}
        \label{fig:TCN-w-TA-NMSE}
    \end{subfigure}
    \caption{BER and NMSE simulation results for the VTV-SDWW vehicular channel model.}
    \label{fig:with-or-without-TA}
\end{figure}
\subsection{Bit Error Rate (BER)}
Figure \ref{fig:TCN-no-TA-BER} shows the BER of the proposed TCN-DPA with and without TA compared to the classical methods, which are DPA, LS, STA, CDP, TRFI and DPA-TA. We also compare the performance of a TCN without any post-processing which is our baseline TCN to determine the need for DPA as a post-processing method. The baseline TCN performance is poor compared to other estimators. The TCN-based estimators clearly showed better performance than the classical estimators. It is observable that from SNR larger than 20 dB the TCN-DPA estimator starts to outperform the classical estimators, and at 40 dB TCN-DPA obtained BER performance close to 0.7 orders of magnitude lower than TRFI. TCN-DPA-TA outperformed all estimators in all SNRs. At 40 dB, the TCN-DPA-TA is approximately 1 order of magnitude lower than the TRFI, which is the best classical estimator.    
The performance of the TCN-DPA estimator at SNR levels less than 20 dB is close to that of classical estimators, regardless of the extensive hyperparameter tuning. This necessitates the addition of TA to the TCN-DPA to smooth the channel estimate per subcarrier by considering the current and previous channel estimate values. The improvement in BER performance of TCN-based estimators can be attributed to their ability to effectively capture relationships between adjacent subcarrier frequency DPA estimates. The TCN learns how the channel affects adjacent subcarriers, and using this information the TCN can accurately predict the channel's response for each individual subcarrier. This allows for a more accurate reconstruction of channel state information across the frequency domain. The better BER of TCN-DPA-TA compared to that of TCN-DPA in low SNR regions further demonstrates the effectiveness of TA postprocessing in refining these estimates. 


\subsection{Normalised Mean Squared Error (NMSE)}
Fig.~\ref{fig:TCN-w-TA-NMSE} shows the NMSE performance of the TCN-based estimators compared to the classical methods namely DPA, LS, STA, CDP, TRFI, DPA-TA and a baseline TCN. The TCN-DPA-TA estimator (red line graph) shows even better performance than the TCN-DPA estimator (blue line graph) because of the added temporal averaging process. It is also evident in this figure that the performance of the TCN-DPA estimator at low SNRs is similar to that of the classical estimators, but from 20 dB its performance increases exponentially outperforming all the classical estimators. For the TCN-DPA-TA, it starts outperforming all the classical estimators from 10 dB, which is lower than that of the TCN-DPA.  

\balance
\section{Conclusion}
\label{conclusion}
In this paper, we propose the application of a tuned TCN with DPA and TA for channel estimation in VTV-SDWW TDL vehicular communications using adjacent subcarriers as input to the TCN. Our performance comparisons between TCN-based estimators and classical estimators demonstrate the suitability of TCNs for vehicular channels. We introduce a new method of structuring data for NNs, by separating and interleaving the real and imaginary components of complex data along the OFDM symbol dimension. Our experiments show that the BER performance of the TCN-based estimators achieves improvements of up to one order of magnitude for TCN-DPA-TA and up to 0.7 orders of magnitude for TCN-DPA. These results demonstrate the superior channel estimation capabilities of our proposed estimators in vehicular communications. In the future, we plan to conduct a detailed computational complexity analysis of the proposed estimators and investigate them in other vehicular channel models.

\section*{Acknowledgment}
The authors gratefully acknowledge the financial support
of this study by the Telkom CoE at NWU and NITheCS. 

 \bibliographystyle{splncs04}
 \clearpage
 \bibliography{references}

\end{document}